\documentclass{PoS}
\usepackage{graphicx}
\usepackage{subcaption}
\usepackage{braket,amsmath,hyphenat,textcomp}
\usepackage[numbers,sort&compress]{natbib}

\title{First calculation of $\hat{q}$ on a quenched SU(3) plasma }

\ShortTitle{First calculation of $\hat{q}$ on a quenched SU(3) plasma}

\author{\speaker{Amit Kumar}\\
        Department of physics and astronomy, Wayne State University, Detroit, MI, USA\\
        E-mail: \email{kumar.amit@wayne.edu}}
      
     \author{Abhijit Majumder\\
        Department of physics and  astronomy, Wayne State University, Detroit, MI, USA\\
       E-mail: \email{majumder@wayne.edu}}

\author{Chiho Nonaka\\
        Department of physics, Nagoya University, Japan \\
       E-mail: \email{  nonaka@hken.phys.nagoya-u.ac.jp }}

\abstract{The jet transport coefficient $\hat{q}$ is the leading transport coefficient that controls the modification of hard jets produced in heavy\hyp ion collisions. This coefficient is inherently non\hyp perturbative, and hence, is challenging to compute from first principles. In this report, we present a perturbative quantum chromodynamics (pQCD) and lattice gauge theory based formulation to study $\hat{q}$. We formulate $\hat{q}$ within a 4\hyp dimensional (4D) quenched SU(3) lattice. We consider a leading order diagram for a hard parton passing through the quark\hyp gluon plasma. The non\hyp perturbative part is expressed in terms of a non\hyp local (two\hyp point) Field\hyp Strength\hyp  Field\hyp Strength (FF) operator product which can be Taylor expanded after analytic continuation to the Euclidean region. Such an expansion allows us to write $\hat{q}$ in terms of the expectation of local operators. Finally, we present our results for $\hat{q}$ in a pure gluon plasma.  }

\FullConference{The 36th Annual International Symposium on Lattice Field Theory - LATTICE2018\\
		22-28 July, 2018\\
		Michigan State University, East Lansing, Michigan, USA.}

\begin{document}
\section{Introduction}
Over the past decades, the phenomenon of jet quenching  has been well established as an indicator of the formation of the quark-gluon plasma (QGP) in heavy-ion collisions.  Among existing known coefficients characterizing transport properties of the QGP, the jet transport coefficient $\hat{q}$ is the leading transport coefficient that controls the modification of jets inside the QGP. 
The transport coefficient $\hat{q}$ is defined as average squared transverse momentum broadening per unit length of the medium. Previously, several methods to compute  $\hat{q}$ from first principles have been attempted, each with its own assumptions,  limitations, and region of validity \cite{ADSCFSTcalculation1234, 
HTLcalculation,
WilsonLineApproachCalculation, KRajagopal2013,  
QuenchedSU2Calculation,
MPanero2014,    
LaineAndRothkopf2014,
BurkeEtal,
AKumarJetPuzzle}.
 A finite-temperature calculation based on Hard Thermal Loop (HTL) predicts $\hat{q}$ to scale as a product of $T^3$ times $\mathrm{log}(E/T)$ \cite{HTLcalculation}. 
A lattice gauge theory based approach has also been  put forward by one of the authors   \cite{QuenchedSU2Calculation}  to compute $\hat{q}$ on a 4D quenched SU(2) plasma.
A well known state-of-the-art phenomenological extraction of $\hat{q}$ has come from the work by the JET collaboration \cite{BurkeEtal}. This extraction is based on parameterizing $\hat{q}$ as a dimensionless constant times $T^3$ and comparison of full model calculations to the experimental data for the nuclear modification factor ($R_{AA}$) of leading hadrons in central collisions at the Relativistic Heavy-Ion Collider (RHIC) and at the Large Hadron Collider (LHC). 
Beyond this, it is possible that $\hat{q}$ possesses a dependence on the resolution scale of the jet \cite{AKumarJetPuzzle}.

In this paper, we report on the  calculation of the transport parameter $\hat{q}$ for a pure gluon plasma extracted using 4D lattice gauge theory. We follow the methodology  described in the article \cite{QuenchedSU2Calculation} and outline  a method to compute $\hat{q}$ from first principles applicable to a hot quark-gluon plasma.

\section{Transport coefficient $\hat{q}$ for a hot QGP}
The transport parameter $\hat{q}$ essentially measures the transverse scattering experienced by a projectile passing through the plasma. 
A framework to evaluate $\hat{q}$ from first principles using lattice gauge theory was first proposed in Ref. \cite{QuenchedSU2Calculation}. In this section, we briefly discuss the {\it{ab-initio}} formulation of $\hat{q}$. Consider the propagation of a hard virtual quark with light-cone 
momentum $q=(\mu^2/2q^{-},q^{-}, 0_{\perp})\sim (\lambda^{2}Q , Q , 0)$ through a section of hot quark-gluon plasma at temperature $T$, where, $ \lambda << 1 , Q >>  \Lambda_{\mathrm{QCD}} $, and $\mu$ is off-shellness of the hard quark. We consider a leading order process of the hard quark traveling along the negative $z$-direction, exchanging a transverse gluon with the plasma. We consider this process in the rest frame of the medium with the momentum of the exchanged gluon as $k=(k^{+}, k^{-}, k_{\perp}) \sim (\lambda^{2} Q, \lambda^{2}Q,\lambda Q)$.
Applying standard pQCD techniques to this leading process, one obtains the following expression for $\hat{q}$,
\begin{equation}
\hat{q} = \frac{ \langle \vec{k}^{2}_{\perp} \rangle }{L}=\frac{8 \sqrt{2} \pi \alpha_{s} }{N_c}    \int \frac{ dy^{-} d^{2} y_{\perp} } {(2\pi)^3}  d^{2} k_{\perp}  e^{ -i  \frac{\vec{k}^{2}_{\perp}}{2q^{-}} y^{-} +i\vec{k}_{\perp}.\vec{y}_{\perp} } \sum _{n}  \bra{n} \frac{e^{-\beta E_{n}}}{Z} F^{+ \perp _{\mu}}(0) F^{+}_{\perp_{\mu}}(y^{-},y_{\perp}) \ket{n},
\end{equation}
 where $L$ is the length of the box, $F^{ \mu\nu} = t^{ a} F^{ a\mu\nu}$ is the gauge field strength, $\alpha_{s}$ is the strong coupling constant, $\beta$ is the inverse temperature, $\ket{n}$ is a state with energy $E_n$, $Z$ is the  partition function of the thermal medium, and $N_{c}$ is the number of colors. Computing the thermal expectation value of the operator $ F^{+ \perp _{\mu}}(0) F^{+}_{\perp_{\mu}}(y^{-},y_{\perp})$ is challenging due to the light-cone separation of the two operators. To turn the expression of $\hat{q}$ into a series of local operators, we define a generalized coefficient as 
\begin{equation}
\hat{Q}(q^{+}) = \frac{ 16\sqrt{2}\pi \alpha_{s}  }{N_c} \int \frac{ d^{4}y d^{4}k }{(2\pi)^4} e^{iky} q^{-}  \frac{ \bra{M}  F^{+ \perp _{\mu}}(0) F^{+}_{\perp_{\mu}}(y) \ket{M}  }{(q+k)^2 + i\epsilon },
\end{equation} 
 where  $\ket{M} $ represents the state of the thermal medium. We do an analytic continuation of $\hat{Q}$ in a $q^{+}$ complex plane. We note that the $\hat{Q}$ has a branch cut due to the quark propagator having momentum $q+k$ in a region where $q^{+} \sim T $. Moreover, we can show that 
\begin{equation}
\left. \frac{ Disc[\hat{Q}(q^{+})] }{2\pi i} \large \right | _{ \hspace{1mm } \mathrm{at}  \hspace{1mm} q^{+} \sim T } = \hat{q}.
\label{eq:qhatDiscontinuity}
\end{equation}
We also note that there is an additional discontinuity in the region $q^{+} \in (0, \infty )$ due to vacuum-like radiative processes. However, when one takes $q^{+} <<0$, say $q^{+}=-q^{-}$, one can expand the quark propagator as follows:
\begin{equation}
\frac{1}{(q+k)^2} \simeq \frac{1}{2q^{-}(-q^{-} + (k^{+} - k^{-}))} = - \frac{1}{2 (q^{-})^2} \left [  \sum^{\infty} _{n=0}  \left ( \frac{\sqrt{2}k_{z}}{q^{-}} \right )^n \right ].
\end{equation}
For this case, we may replace the gluon momentum $k_{z}$ with regular derivative $\partial_{z}$ acting on the Field-Strength $F^{+}_{\perp_{\mu}}(y)$. On adding the contributions from gluon scattering diagrams, we enhance the regular derivative into a covariant derivative. Thus, we arrive at
\begin{equation}
\hat{Q}( q^{+}=-q^{-}) =  \frac{8\sqrt{2} \pi \alpha_{s}  }{N_c q^{-}} \bra{M} F^{+ \perp_{\mu}}(0) \sum^{\infty}_{n=0}   \left(   \frac{i\sqrt{2}D_{z} }{q^{-}}\right)^{n} F^{+}_{\perp_{\mu}}(0)   \ket{M}.
\end{equation} 
In the above equation, each term in the series is local, and hence, one can directly compute their expectation value on the thermal lattice. Note that $\hat{Q}(q^{+}=-q^{-})$ is not the transport coefficient $\hat{q}$, however, both are related. Consider the following contour integral in the $q^{+}$ complex plane:
\begin{equation}
I_{1} = \oint \frac{dq^{+}}{2\pi i} \frac{\hat{Q}(q^{+})}{(q^{+}+q^{-})},
\end{equation}
where the contour is taken as a counter-clockwise circle centered around point $q^{+}=-q^{-}$ and with a radius small enough to exclude regions where $\hat{Q}(q^{+})$ is discontinuous. To evaluate this integral, we deform the contour and evaluate the integral over the branch cut $q^{+} \in (-  T, \infty)$:
\begin{equation}
\hat{Q}(q^{+} = - q^{-}) = \int ^{T_2} _{-T_1} \frac{dq^{+}}{2\pi i} \frac{Disc[\hat{Q}(q^{+})]}{q^{+} + q^{-} }  + \int ^{\infty}_{0} \frac{dq^{+}}{2\pi i} \frac{Disc[ \hat{Q}(q^{+})]}{(q^{+} + q^{-})},
\label{eq:ContourEquationThermalPlusVacuum}
\end{equation}
 where $T_{1}+T_{2} \sim T$ represents a width of the thermal discontinuity of $\hat{Q}(q^{+})$ in $q^{+}$ real axis. The first integral in  Eq. \ref{eq:ContourEquationThermalPlusVacuum} represents the contribution from the interaction of the hard incident quark with the medium. Note that $q^{+} \in [-T_1,T_2]$ represents a region of the thermal discontinuity in $\hat{Q}(q^{+})$. The second integral in Eq. \ref{eq:ContourEquationThermalPlusVacuum} represents the contribution from a vacuum-like processess, where the hard quark with momentum $q^{+} \in (0, \infty)$ is time-like and undergoes vacuum-like splitting. Thus, this second integral is temperature independent. 
At this point, we employ Eq.  \ref{eq:qhatDiscontinuity} to arrive at an expression for the average $\hat{q}$ given as
\begin{equation}
\hat{q} = \frac{ 8\sqrt{2}\pi \alpha_{s}  }{N_c (T_{1} + T_{2})} \bra{M} F^{+ \perp_{\mu}}(0) \sum^{\infty}_{n=0}   \left(   \frac{i\sqrt{2}D_{z} }{q^{-}}\right)^{n} F^{+}_{\perp_{\mu}}(0)    \ket{M}_{(\mathrm{Thermal-Vacuum})}.
\label{eq:qhatLatticeEquation}
\end{equation} 
The above expression is a desired form of transport coefficient $\hat{q}$ suitable for lattice calculation which contains several features. First, each of the terms in the series are local, that means one can hope to compute their expectation value on the thermal lattice. Second, higher order terms in the series are suppressed by the hard scale $q^{-}$, and hence, computing first few terms may be sufficient. Also, we emphasize that we have not made any assumptions regarding the constituents of the plasma, and hence, the expression of the transport coefficient $\hat{q}$ given in Eq.  \ref{eq:qhatLatticeEquation} is valid for both pure gluonic thermalized plasma and full quark-gluon thermalized plasma. It is also interesting to mention that a similar kind of operator product expansion has been found by the author of Ref. \cite{XJiPartonPDF} in  his analysis of the parton distribution function on an Euclidean space.
In the next section, we will discuss the evaluation of the local operators on the lattice.
\section{Computing local operators on the lattice}
In our first attempt, we have computed $\hat{q}$ by considering the first two terms in the series (Eq.  \ref{eq:qhatLatticeEquation}) and have ignored higher order terms.
The correlators to be evaluated are $\bra{M} F^{+\perp_{\mu}}(0) F^{+}_{\perp_{\mu}}(0)\ket{M} $ and $\bra{M} F^{+ \perp_{\mu}}(0) \left(  \frac{i\sqrt{2}D_{z}}{q^{-}}  \right)  F^{+}_{ \perp_{\mu}}(0) \ket{M}$. But, we note that these operators are in Minkowski space. In order to compute their expectation value on the lattice, we rotate the operator products to Euclidean space. This is achieved by following transformations:
\begin{equation}
x^{0} \longrightarrow -i x^{4},
 \hspace{2mm} A^{0} \longrightarrow  iA^{4} \Longrightarrow  F^{0i} \longrightarrow iF^{4i}.
\end{equation}
This leads to 
\begin{equation}
F^{+ \perp_{\mu}}(0) F^{+}_{ \perp_{\mu}}(0)  \overset{}{\longrightarrow} \frac{1}{2}  \left[    \sum^{2}_{i=1}( F^{3i}F^{3i} - F^{4i}F^{4i})    + i \sum^{2}_{i=1} (F^{4i}F^{3i} + F^{3i}F^{4i})   \right]
\label{LOoperators}
\end{equation}
and 
\begin{equation}
\begin{split}
 F^{+ \perp_{\mu}} \left(  \frac{i\sqrt{2}D_{z}}{q^{-}}  \right)  F^{+}_{ \perp_{\mu}}  
  \overset{}{\longrightarrow}   \frac{ \sqrt{2}}{2q^-}   \left[  i  \sum^{2}_{i=1}( F^{3i} D_{z} F^{3i} - F^{4i} D_{z} F^{4i})    -   \sum^{2}_{i=1} (F^{4i} D_{z} F^{3i} + F^{3i} D_{z}  F^{4i})  \right].
 \label{LOoperatorsWithDzDerivatives}
\end{split}
\end{equation}
Now, we set up a four-dimensional (4D) grid, specified by the coordinate $x_\mu = a_{L} * n_{\mu}$, where
$n_\mu = (n_x, n_y, n_z,  n_\tau)$ is a 4-component Euclidean vector. Here $n_x$, $n_y$, $n_z$,  $n_\tau$  represents the number of grid  points in $x$, $y$, $z$, and   $\tau$ direction. We denote $a_{L}$ to represent the lattice spacing. Then, the temperature is given by 
$T=1/n_{\tau}a_{L}$.
In our calculations, lattices with the same lattice spacing in all directions are employed. We consider the lattice to be symmetric  in the spatial directions, i.e. $n_x=n_y= n_z=n_{s}$.  For calculations at finite temperature, the 
number of sites in the spatial direction ($n_s$) is set to be a multiple of number of sites in the temporal direction ($n_{\tau}$), whereas vacuum calculations are done with the same number of sites in all four directions. We use a heat bath algorithm to generate our gauge field configurations.
 We evaluate the Euclidean operators in Eq. \ref{LOoperators} and \ref{LOoperatorsWithDzDerivatives}   by expressing field-strength operator and its covariant derivatives in terms of plaquette variables on the lattice as
\begin{equation}
F^{\mu\nu} (x)= t^{a} F^{a \mu\nu}(x) =\frac{ U^{\mu\nu}(x) - U^{ \dagger \mu \nu } (x)}{2iga^{2}_{L}} ,  \hspace{4mm}
D_z F^{\mu\nu}(x) =\frac{ F^{\mu\nu}(x_{3}
+a_{L}) - U_{3}(x)F^{\mu\nu}(x) }{ a_{L}},
\end{equation}
where $U^{\mu\nu}$ is a plaquette in the $\mu\nu$ plane, and $ g$ is the bare coupling constant.
\begin{figure}[h!]
  \centering
  \begin{subfigure}[b]{0.48\linewidth}
    \includegraphics[width=\linewidth]{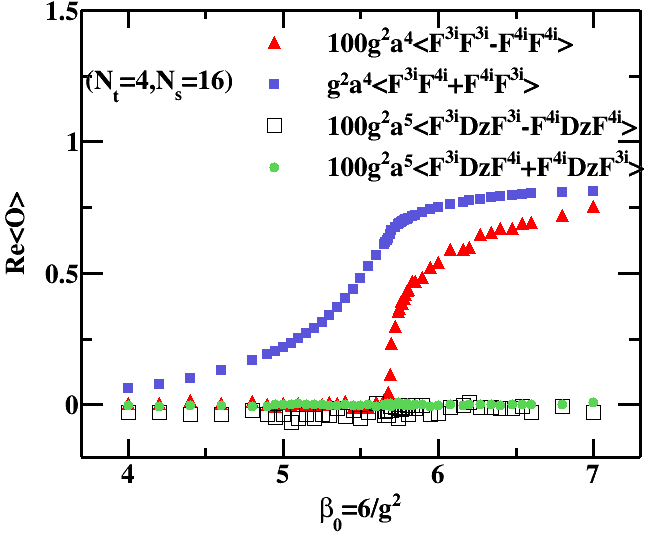}
    \caption{}
  \end{subfigure}
  \begin{subfigure}[b]{0.48\linewidth}
    \includegraphics[width=\linewidth]{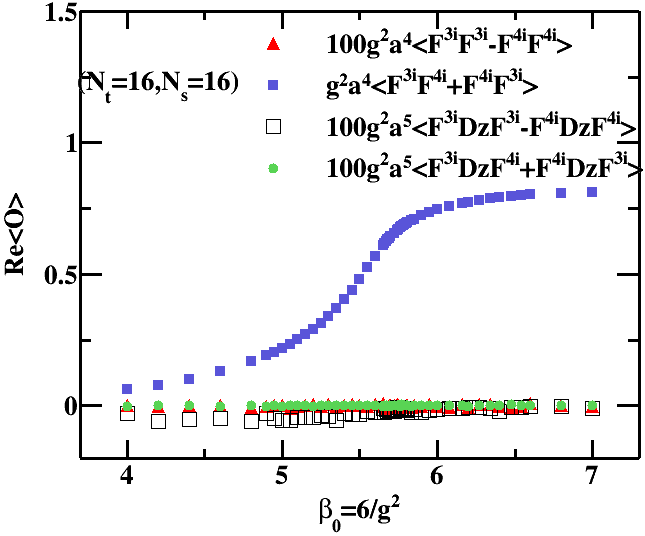}
    \caption{}
  \end{subfigure}
  \caption{Expectation value of the operators on quenched $SU(3)$ plasma as a function of the bare coupling constant for $n_{\tau}=4$, $n_{s}$=16.  (a) Thermal + Vacuum contribution. (b) Vacuum contribution. }
  \label{fig:OperatorsWrtBareCouplingConstant}
\end{figure}

In this first attempt, we have evaluated operators for the quenched case, and left the unquenched calculation for future work. In order to generate gauge field configuration for quenched calculation, we used the Wilson gauge action for pure $SU(3)$ gauge field.  All the calculations have been done by taking a statistical average over 5000 gauge
configuration generated using the standard  heat-bath algorithm. 
We have computed operators for $n_{\tau}=$ 4, 6, 8 with $n_{s}=4n_{\tau}$ 
as function of input parameter $\beta_{0}=6/g^{2}$. In Fig. \ref{fig:OperatorsWrtBareCouplingConstant}, we show the expectation value of operators as a function of $\beta_{0}$ for the lattice size $n_{\tau}=4$ and corresponding  vacuum contribution. The uncrossed correlator $\langle F^{3i}F^{3i}-F^{4i}F^{4i} \rangle $ shows a rapid transition in the region $\beta_{0} \in (5.5, 6.0)$ and has negligible contributions from vacuum processes compared to the pure thermal part. The crossed correlator $\langle  F^{3i}F^{4i} + F^{4i}F^{3i} \rangle $ shows a smooth transition in the region $\beta_{0} \in (5.0, 6.0)$ and has vacuum contributions as dominant  compared the pure thermal part. 
However, the correlators with $D_z$ derivatives   $\langle   F^{3i}D_{z}F^{3i}-F^{4i}D_{z}F^{4i}  \rangle $  and  $\langle   F^{3i}D_{z}F^{4i}+F^{4i}D_{z}F^{3i}  \rangle $  look suppressed for all values of the bare coupling constant. To understand the behavior of these operators in terms of physical quantities, we need to relate the bare coupling constant with the lattice spacing.
This relation is set using the two-loop perturbative renormalization group  (RG) equation with non-perturbative correction \cite{SU3ScaleSettings} given as 
\begin{equation}
a_{L} = \frac{f}{\Lambda_{L}} \left[ \frac{11g^2}{16\pi^2} \right]^{\frac{-51}{121}}  \exp  \left[ \frac{-8\pi^2}{11g^2} \right],
\label{eq:twoLoopBetaFunction}
\end{equation}
 where $\Lambda_{L}$ is a dimensionful parameter, and $f$ is non-perturbative correction. Note, the
temperature  $T$ is obtained by $1/(n_{\tau} a_{L})$. 
\begin{figure}[h]
  \centering
  \begin{subfigure}[b]{0.49\linewidth}
    \includegraphics[width=\linewidth]{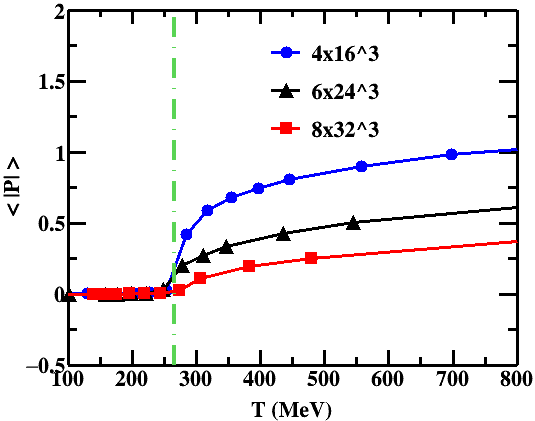}
    \caption{}
  \end{subfigure}
  \begin{subfigure}[b]{0.49\linewidth}
    \includegraphics[width=\linewidth]{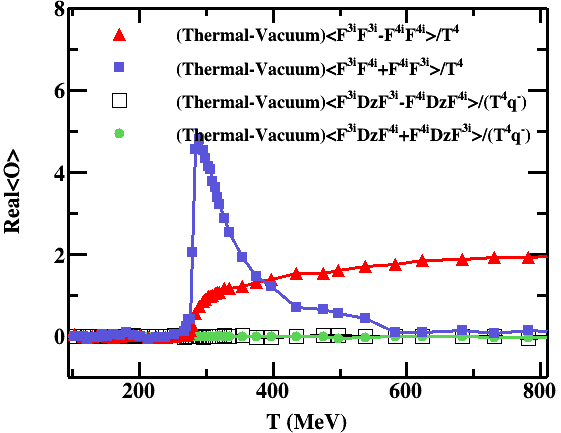}
    \caption{}
  \end{subfigure}
  
   \begin{subfigure}[b]{0.49\linewidth}
    \includegraphics[width=\linewidth]{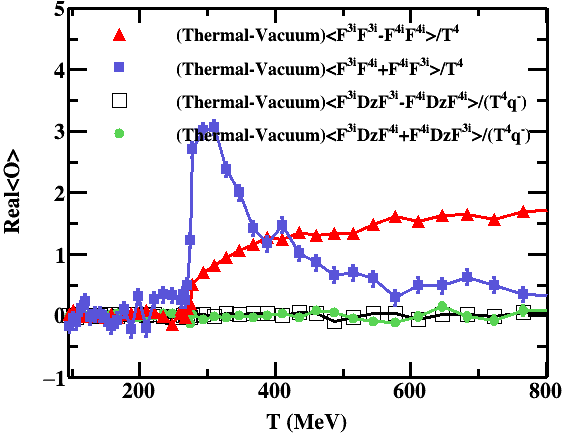}
    \caption{}
  \end{subfigure}
  \begin{subfigure}[b]{0.49\linewidth}
    \includegraphics[width=\linewidth]{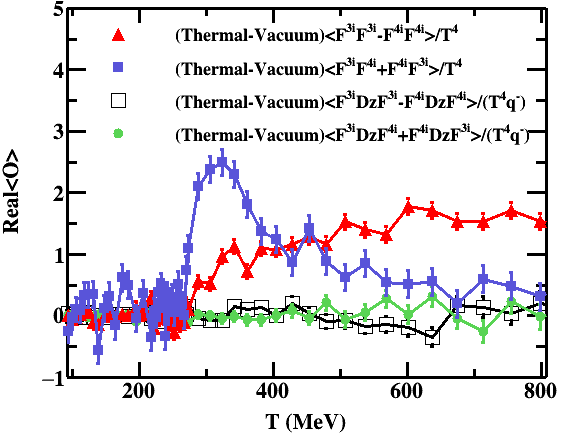}
    \caption{}
  \end{subfigure}
  \caption{Expectation value of operators as a function of temperature for pure gluon plasma. We used $q^{-}$ as 20 GeV. (a) The Polyakov loop for $n_{\tau}=$ 4, 6, and 8.  (b)  The real part of FF correlators for $n_{\tau}=4$ and $n_{s}=16$. (c) The real part of FF correlators for $n_{\tau}=6$ and $n_{s}=24$. (d) The real part of FF correlators for $n_{\tau}=8$ and $n_{s}=32$.}
  \label{fig:OperatorsWrtTemperature}
\end{figure}
To estimate the non-perturbative correction factor $f$, we evaluate the average value of the Polyakov loop ($\langle  |P| \rangle $) and adjust the free parameter $f$ such that $T_{c}/\Lambda_{L}$ is independent of bare coupling constant $g$, where $T_{c}\sim 265 $ MeV is the critical temperature for pure $SU(3)$ gauge theory \cite{SU3ScaleSettings}. 
The value of the Polyakov loop ($P$) for a given gauge configuration is given as
\begin{equation}
P = \frac{1}{n_{x}n_{y}n_{z}}tr \left[  \sum_{ \vec{r}} \prod ^{n_{\tau}-1}_{n=0} U_{4}(na,\vec{r})   \right],
\end{equation} 
where $U_{4}(na,\vec{r})$ is a gauge link in the temporal ($\tau$) direction. 
In  Fig. \ref{fig:OperatorsWrtTemperature}(a), we show the expectation value of the Polyakov loop $\langle  |P| \rangle $  as a function of temperature for $n_{\tau}=$ 4, 6 and 8 after performing the non-perturbative renormalization. The green vertical line represents the critical temperature for pure $SU(3)$ gauge theory (see Fig. \ref{fig:OperatorsWrtTemperature}(a)).

We present the expectation value of operators as a function of temperature in Figs. \ref{fig:OperatorsWrtTemperature}(b), \ref{fig:OperatorsWrtTemperature}(c) and \ref{fig:OperatorsWrtTemperature}(d) for $n_{\tau}=$ 4, 6 and 8, respectively. The operators plotted are scaled by $T^{4}$ or $T^4q^{-}$ to make them dimensionless. We observe that the uncrossed correlator  $\langle   F^{3i}F^{3i}-F^{4i}F^{4i} \rangle $ (red curve) shows a rapid transition near the critical temperature $T\in (250,350)$ MeV and is dominant at high temperature compared to rest of the operators (see Figs. \ref{fig:OperatorsWrtTemperature}(b,c,d)). The crossed correlator   $\langle  F^{3i}F^{4i}+F^{4i}F^{3i} \rangle $ (blue curve) shows a peak-like behavior near the transition region $T \in (250,350)$ MeV and is suppressed at high temperature (see Figs. \ref{fig:OperatorsWrtTemperature}(b,c,d)). We also observe that both the operators with covariant derivatives (green and black curves) are suppressed at all temperatures (see Figs. \ref{fig:OperatorsWrtTemperature}(b,c,d)). 

\section{Results and discussions}
\begin{figure}[h!]
  \centering
  \begin{subfigure}[b]{0.45\linewidth}
    \includegraphics[width=\linewidth]{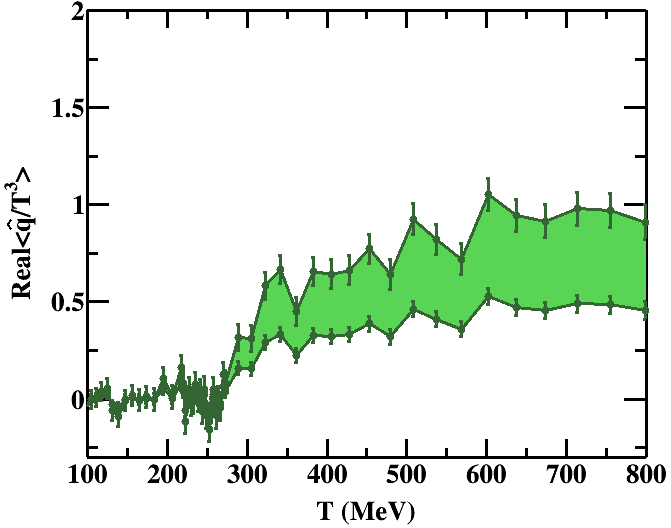}
    \caption{}
  \end{subfigure}
  \begin{subfigure}[b]{0.45\linewidth}
    \includegraphics[width=\linewidth]{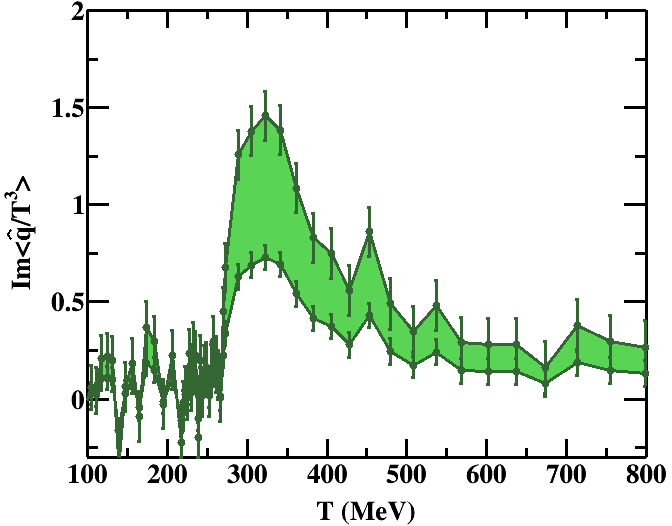}
    \caption{}
  \end{subfigure}
  \caption{$\hat{q}/T^3$ as a function of temperature for hard quark propagating through a pure gluonic plasma. (a) The real part of $\hat{q}/T^3$ w.r.t $T$. (b) The Imaginary part of $\hat{q}/T^3$ w.r.t $T$.}
  \label{fig:qhatT3WrtT}
\end{figure}

We show in Fig. \ref{fig:qhatT3WrtT} our final result of $\hat{q}$ computed on the quenched $SU(3)$ lattice. 
The figure \ref{fig:qhatT3WrtT}(a) represents the real part of $\hat{q}/T^3$ as a function of temperature. It exhibits a transition region $T\in (250,350)$ MeV. At high temperature, the extracted value of $\hat{q}/T^3$ turns out to be $\sim 0.5-1$. We have also plotted the imaginary part of $\hat{q}/T^3$ as a function of temperature in Fig. \ref{fig:qhatT3WrtT}(b). Naively, one expects it to be zero, but our calculation shows that it has a peak-like behavior near the transition region $T\in (250,350)$ MeV, and goes to zero at high temperature. The contribution to imaginary part of $\hat{q}/T^3$ comes from the fact that the vacuum subtracted expectation value of cross-correlator $\langle  F^{3i}F^{4i}+F^{4i}F^{3i}  \rangle $ is non-zero near the transition region. The physical interpretation of this crossed-correlator in terms of chormo-elctro-magnetic field is that it represents $(\vec{E} \times \vec{B})_{z} $, a Poynting vector in the $z$-direction, causes damping effects for thermal gluons emerging from the plasma. We also point out that the band in Fig. \ref{fig:qhatT3WrtT} originates from the lack of precise knowledge of the width of the thermal discontinuity ($T_{1}+T_{2}$) in the  generalized object $\hat{Q}(q^{+})$ in the $q^{+}$ complex plane. We used a hard thermal loop calculation to estimate this, where $T_{1}+T_{2} \sim 2T-4T$.

\section{Summary and future work}
In this work, we have established a first principles framework of $\hat{q}$, applicable to 4D hot quark-gluon plasma. We computed for the first time, the transport coefficient $\hat{q}$ for a pure gluon plasma using lattice gauge theory. We considered a leading order process where the hard quark produced from the hard scattering in heavy-ion collisions propagates through the hot plasma while exchanging a transverse gluon with the plasma. In order to express $\hat{q}$ in terms of local operators, we defined a generalized coefficient $\hat{Q}(q^{+})$ on the $q^{+}$ complex plane and showed how it is connected to the physical $\hat{q}$ when studied in the region where $|q^{+}| << q^{-}$. We also found that this object in region $q^{+} \lesssim -q^{-}$, can be expressed in terms of a  series of local operators. We used the method of dispersion relations to relate the two regions. In this study, we evaluated operators on a quenched $SU(3)$ plasma for different lattice sizes, and found that the results displayed scaling behavior. Our quenched lattice result constrains $\hat{q}/T^3 \sim 0.5 -1$ at high temperatures. We also found that the thermal gluons emerging from the plasma could undergo a damping process giving   a non-zero peak-like structure in imaginary part of $\hat{q}/T^3$ in the transition region.

The lattice formulation of $\hat{q}$ discussed in this paper marks a crucial step towards realizing a true {\it{ab-initio}} formulation of $\hat{q}$. In future attempts, we aim to extend the evaluation of the local operators on the unquenched $SU(3)$ lattice. We would like to emphasize that the expression of $\hat{q}$ appearing in the Eq. \ref{eq:qhatLatticeEquation}  is valid for both pure gluon plasma and the quark-gluon plasma. 
For future attempts, we will also include the contributions from the medium-induced radiative splitting. To improve the numerical accuracy, we would like to go beyond conventional Wilson\textquotesingle s action to more improved actions.

\section{Acknowledgment}
This work was supported in parts by the National Science Foundation (NSF) under the grant number ACI-1550300 , and also supported by US Department of energy (DOE), office of science, office of nuclear physics under grant number DE-SC0013460.


\begin{thebibliography}{99}

\bibitem{ADSCFSTcalculation1234}
H. Liu, K. Rajagopal and U. A. Wiedemann, Phys. Rev. Lett. 97 182301 
(2006);
F. Lin and T. Matsuo,  Phys. Lett. B641 45-49  (2006);
S. D. Avramis and K. Sfetsos,  JHEP 0701  065  (2007);
N. Armesto, J. D. Edelstein and J. Mas, JHEP 0609 039 (2006).
\bibitem{HTLcalculation}
S. Caron-Huot and C. Gale, Phys. Rev. C82, 064902
(2010).
\bibitem{WilsonLineApproachCalculation}
M. Benzke, N. Brambilla and M. A. Escobedo, A. Vairo,  JHEP 1302  129  (2013).
\bibitem{KRajagopal2013}
F. Eramo, M. Lekaveckas, H. Liu, K. Rajagopal, JHEP 1305 031 (2013). 
\bibitem{QuenchedSU2Calculation}
A. Majumder, Phys. Rev. C87  034905 (2013).
\bibitem{MPanero2014}
M. Panero, K. Rummukainen, A. Schafer, Phys. Rev. Lett. 112 162001 (2014).
\bibitem{LaineAndRothkopf2014}
M. Laine, A. Rothkopf, PoS LATTICE2013  174 (2014); arXiv:1310.2413



\bibitem{BurkeEtal}
K. M. Burke et al. (JET collaboration), Phys. Rev. C90, 014909
(2014).
\bibitem{AKumarJetPuzzle}
A. Kumar, E. Bianchi, J. Elledge, A. Majumder, G. Qin and C. Shen,
Nucl. Phys. A967 536-539  (2017);
A. Majumder $et$ al, arXiv:1702.00481 [nucl-th] (2017).

\bibitem{SU3ScaleSettings}
G. Boyd, J. Engels, F. Karsch, E. Laermann, C. Legeland, M. Lutgemeier and B. Petersson, Nucl. Phys. B469  419-444  (1996);
Owe Philipsen,  Prog. Part. Nucl. Phys. 70 55-107 (2013).

\bibitem{XJiPartonPDF}
X. Ji, Phys.Rev.Lett. 110 262002 (2013).

\end{thebibliography}
\end{document}